\documentclass{aastex61}
\usepackage{graphicx}

\begin{document}
\title{Strong CO$^+$ and N$_2^+$ Emission in Comet C/2016 R2 (Pan-STARRS) }
\correspondingauthor{Anita Cochran}
%\email{anita@astro.as.utexas.edu}
\author{Anita L. Cochran} 
\affil{McDonald Observatory, University of Texas at Austin}
\author{Adam J. McKay}
\affil{NASA Goddard Space Flight Center/Universities Space Research Association }

\footnote{This paper includes data taken at The McDonald Observatory of The University of Texas at Austin}

\begin{abstract}
We report on imaging and spectroscopic observations of comet C/2016~R2 
(Pan-STARRS) obtained with the 0.8\,m and 2.7\,m telescopes of McDonald
Observatory in November and December 2017 respectively. 
The comet was at a heliocentric distance greater than 3\,{\sc {\sc au}}
during both sets
of observations.  The images showed a well-developed tail with properties
that suggested it was an ion tail.  The spectra confirmed that we
were observing well-developed bands of CO$^+$ and N$_2^+$.  The
N$_2^+$ detection was unequivocally cometary and was one of the strongest
bands of N$_2^+$ detected in a comet spectrum.  We derived the
ratio of these two ions and from that we were able to derive that
N$_2$/CO = 0.15.  This is the highest such ratio reported for a comet.
\end{abstract}

\section{Introduction}
Comets represent leftovers from the origins of the Solar System and are an 
amalgam of various ices and dust.  When perturbed into the inner Solar
System, they get heated and the ices sublime forming the coma and any tails.
The most volatile species are sublimed first and are generally exhausted from
near the surface leaving less volatile ices to sublime in subsequent solar
passages.  Spectra of comets consist of emissions from the sublimed ices,
including neutrals and ions, along with a continuum of solar light
reflected off the dust.

Comet C/2016 R2 (Pan-STARRS) (hereafter R2) was discovered by the
Pan-STARRS telescope on 7 September 2016.  With an orbital period of
$\sim 20,550$ years and semi-major axis of $\sim 1500$\,{\sc AU},
this comet came from the Oort cloud, but was not a dynamically new comet.
Its perihelion distance will be at 2.6\,{\sc AU} in May 2018.
In this letter we report on optical imaging and spectroscopic observations of
this comet obtained at The University of Texas McDonald Observatory
in November--December 2017.

\section{Observations}

\begin{figure}[b]
\plottwo{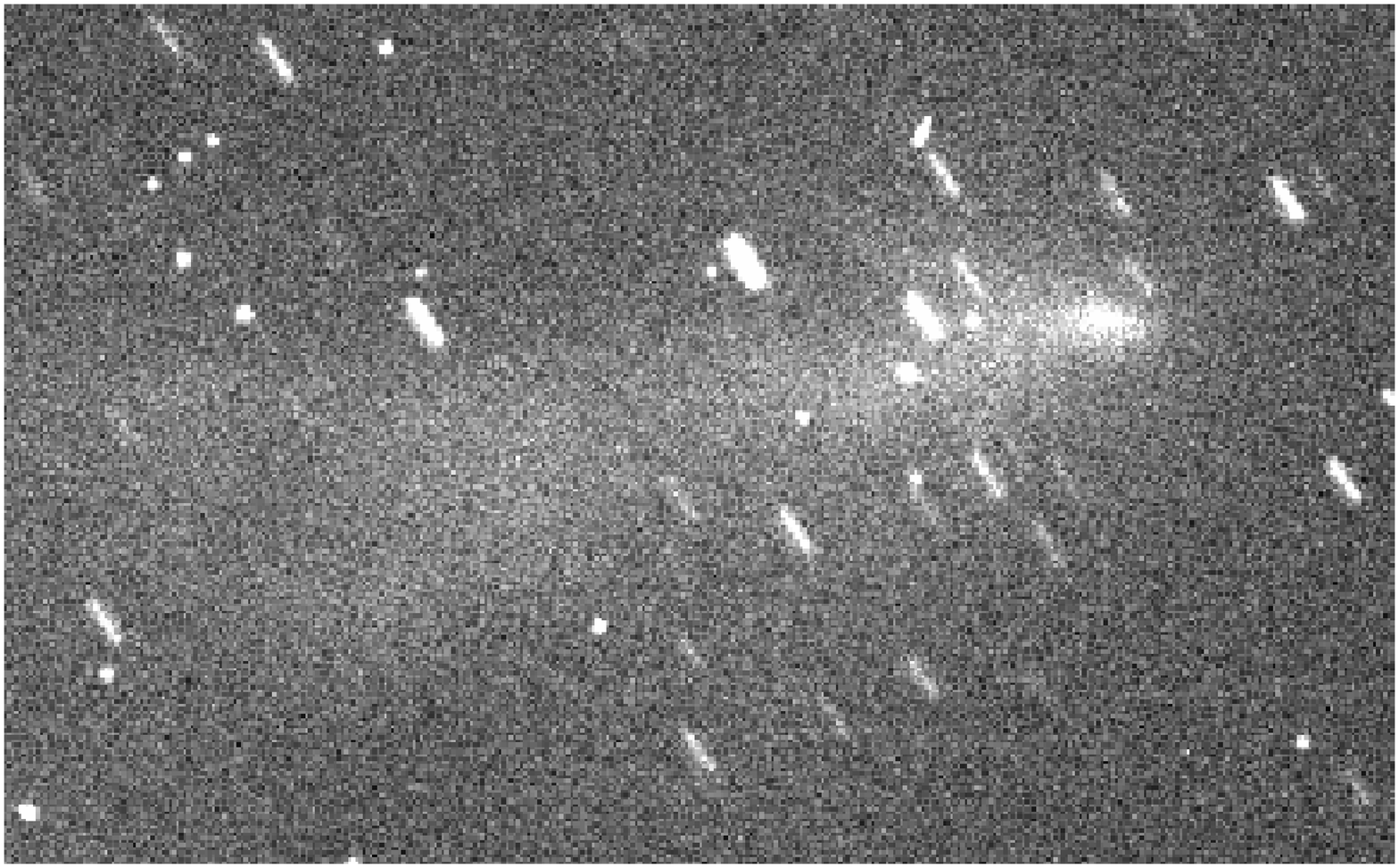}{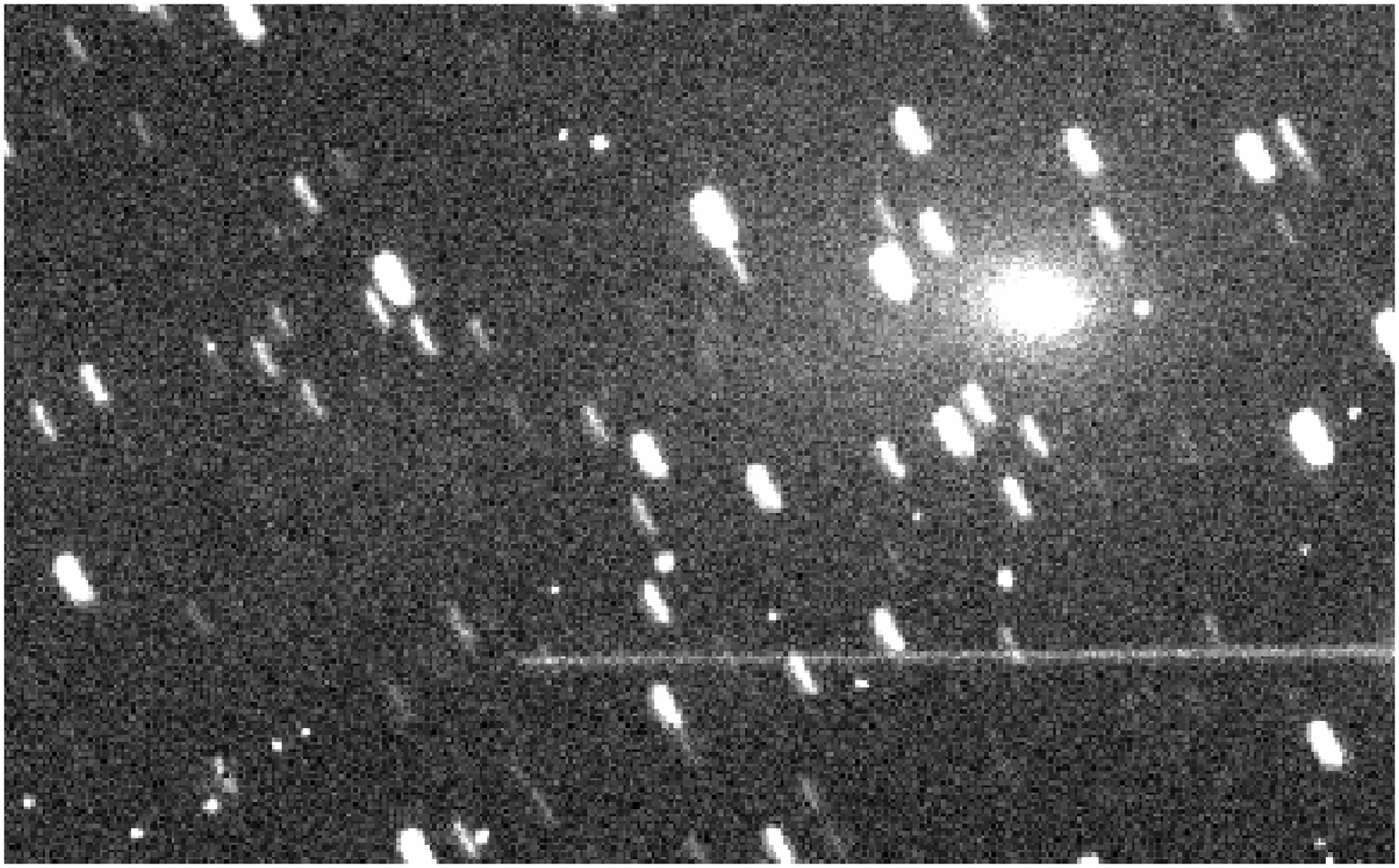}
\caption{Two raw images obtained with the 0.76\,m telescope through the
CN (left) and C$_2$ (right) filters are shown.  Note the tail in both images,
with the tail being much better developed in the CN image.
The images shown here are a subset of the original images and show 
about 10 arcmin on the long axis.
In these images north is down and west is to the left. 
The direction to the
Sun is directly opposite the tail (PA$_{Sun}=120$ degrees).}
\label{image}
\end{figure}

We imaged R2 using the narrow-band Hale-Bopp comet filters
\citep{fascah2000} on  15 November 2017 UT using the Prime
Focus Corrector Camera on the 0.76\,m telescope at McDonald Observatory.
This camera has a 46$\times$46 arcmin$^2$ field-of-view with 1.35\,arcsec
pixels. A log of observations is given in Table~\ref{log}.
Images were obtained with filters intended to isolate emissions of OH, CN,
C$_2$ and two continuum regions in the blue and NUV.  Figure~\ref{image}
shows the images obtained with the CN and C$_2$ filters.  Of note in this
figure is the extremely well developed tail seen in both filters but
substantially stronger and more developed in the CN image.  

It is rare
to see a well developed tail at such a large heliocentric distance.  In
addition, it is extremely unusual to see the tail more developed in the CN
filter.  Our first idea was that this was an ion tail rather than a dust
tail, but even that is uncommon at such a large heliocentric distance.
Consultation with others in the field also suggested that the tail must be ionic
(Schleicher, Farnham, Knight, personal communications, 2017).
In order to confirm our hypothesis of ionic emission contaminating our 
narrow-band imaging, we followed up the images with
spectroscopy.

Spectra were obtained of R2 using the Tull 2DCoude spectrograph 
\citep{TuMQSn95} on
the  Harlan J. Smith 2.7\,m telescope of McDonald Observatory on
8--10 December 2017 UT. Details of the observations are in Table~\ref{log}.
The spectrograph was used with a 1.2 arcsec wide by 8 arcsec tall slit
centered on the optocenter for all observations,
yielding a resolving power ($R=\lambda/\Delta\lambda$) of 60,000. 
The brightness of R2 was around V$\sim12.5$ total.
It was immediately evident
that this comet's spectrum was different than most comets when we read
out the first spectrum on 8 December.  Missing was the strong CN band at
3880~\AA~that is normally one of the strongest emission features observed in
optical spectra of comets, and something we have observed with even
slightly fainter comets when they are closer to the Sun.
Missing also was any
hint of the other usual molecules, C$_2$, C$_3$, CH, or NH$_2$.  Instead
there was a well-developed series of bands scattered from approximately 
3700~\AA~to
5100~\AA.  Most of the bands either degraded to the red or were peaked near
the center of the band.  There was one band that degraded blueward near
3900~\AA. 
\begin{table}
\centering
\caption{Log of Observations} \label{log}
\begin{tabular}{rlrcccc}
\hline
\multicolumn{3}{l}{Images} \\
\multicolumn{3}{l}{Date} & R$_h$ & $\Delta$ & \multicolumn{2}{c}{images per} \\
\multicolumn{3}{l}{(UT)} & (AU) & (AU) & \multicolumn{2}{c}{filter} \\
\hline
15 & Nov & 2017 & 3.19 & 2.32 & \multicolumn{2}{l}{5 C$_2$, 3 CN} \\
\hline
\multicolumn{3}{l}{Spectra} \\
\multicolumn{3}{l}{Date} & R$_h$ & $\Delta$ & $\dot{\Delta}$ & Num. \\
\multicolumn{3}{l}{(UT)} & (AU) & (AU) & (km/sec)$^\dagger$ & (30min) \\
\hline
8 & Dec & 2017 & 3.06 & 2.10 & -10.48 & 1 \\
9 & Dec & 2017 & 3.05 & 2.09 & -9.85 & 6 \\
10 & Dec & 2017 & 3.05 & 2.09 & -9.21 & 6 \\
\hline
\multicolumn{7}{l}{$\dagger$ Doppler shift in December $\sim0.13$\AA} \\
\hline
\end{tabular}
\end{table}

\begin{figure}
\includegraphics[scale=0.66]{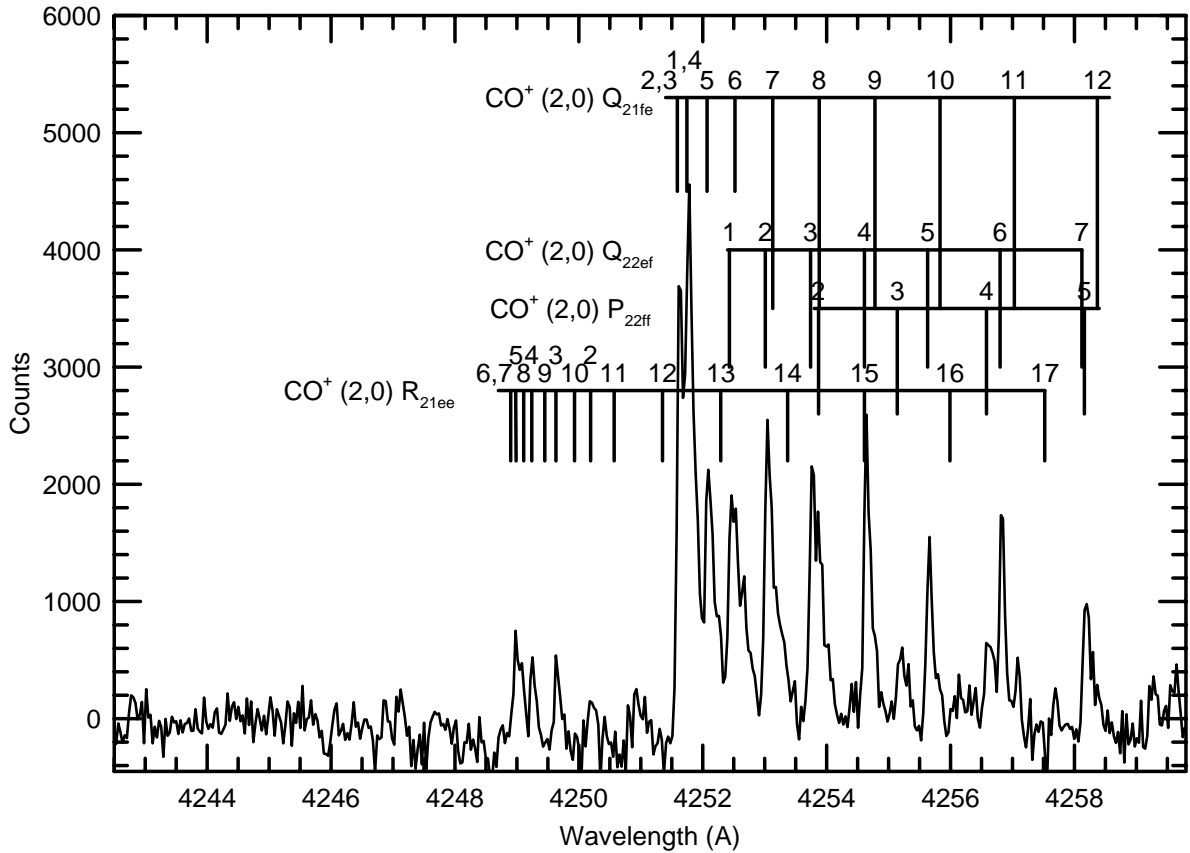}
\caption{The spectrum of the comet at the wavelengths of the $^2\Pi_{1/2}$
ladder of the CO$^+$ (2,0) 
band is shown. The different lines of CO$^+$ are marked. 
The quantum numbers given are $N=J+1/2$, where N is the rotational
quantum number without spin and J is the rotational quantum
number including spin.   The R$_{22ff}$ branch is coincident in wavelength
with the Q$_{21fe}$ branch but with J-value one lower.
This spectrum is the result of averaging all three nights' data after first 
accounting for the Doppler shift on each night.
}\label{coplus}
\end{figure}

The majority of the detected strong bands can be attributed to CO$^+$.  
We
detected the CO$^+$ (4,0), (3,0), (2,0), (1,0), (4,2), (3,2), (2,1) and
(1,1) A~$^2\Pi$ -- X$^2\Sigma$ bands. 
Figure~\ref{coplus} shows the $^2\Pi_{1/2}$ ladder of the CO$^+$ (2,0) band,
though we actually observed both ladders.
As can be seen from the figure, the CO$^+$ bands
are quite complex, with many perturbations and satellite lines.
The wavelengths for these bands come from \citet{kumistmo86}, \citet{haprre92} 
and \citet{haprre00}.

\begin{figure}
\includegraphics[scale=0.66]{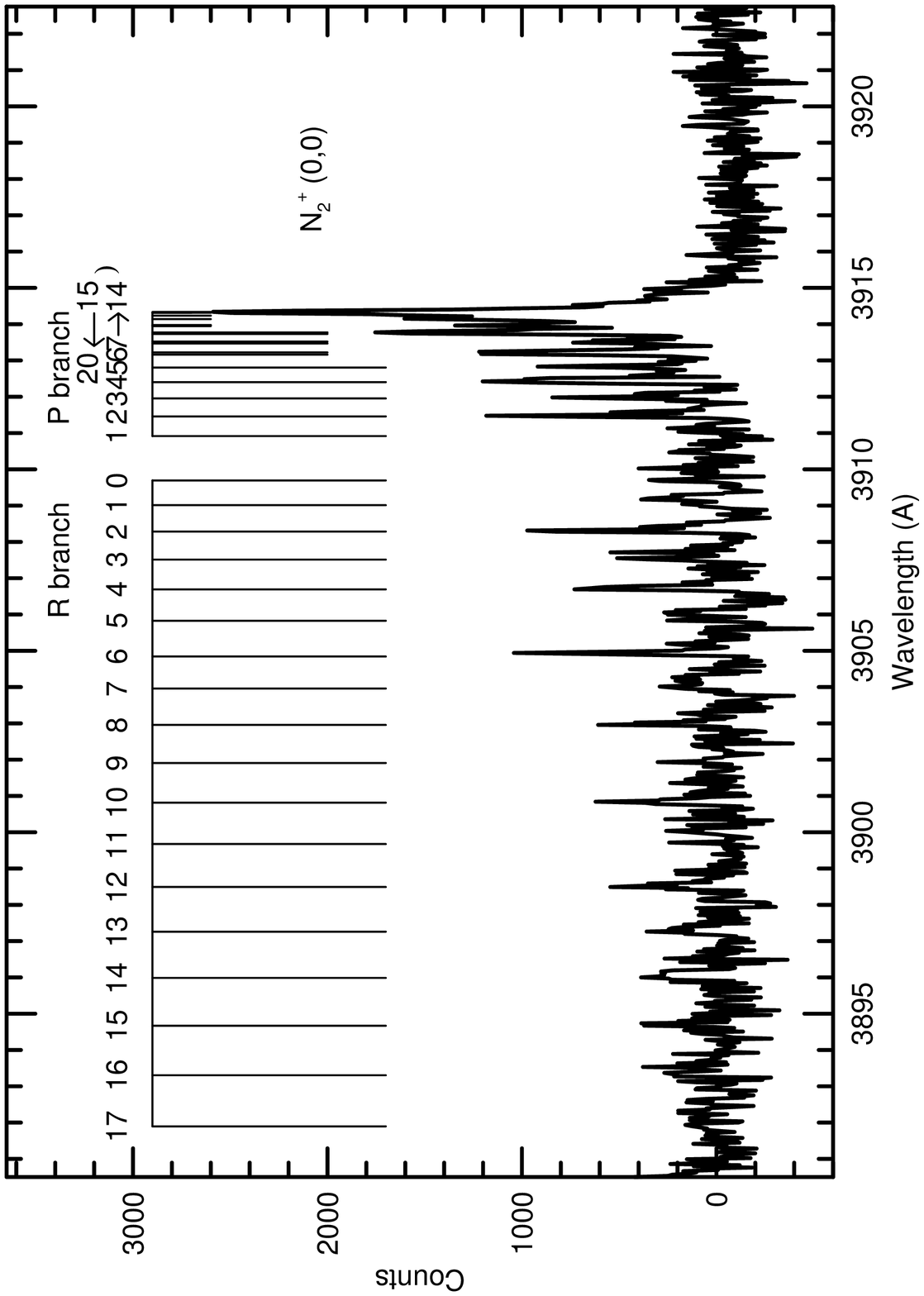}
\caption{The N$_2^+$ region of the spectrum is shown with the P- and R-branch
lines marked.  Clearly, these lines are well detected in this cometary spectrum.
This spectrum is the result of averaging all three nights' data after first 
accounting for the Doppler shift on each night.
} \label{n2plus}
\end{figure}

The blue degrading band can be attributed to the B~$^2\Sigma$ --X~$^2\Sigma$
(0,0) band of N$_2^+$ with a
bandhead at 3914\AA.  Figure~\ref{n2plus} shows this band with the P and R
branches marked (since the band is a $\Sigma-\Sigma$ transition it does not  
have a Q branch).  
The wavelengths and structure of the band comes from \citet{dietal78}.
In this figure note that we see R-branch lines up through J=16
(and quite possibly to J=18). Note also, that the odd J-level R-branch
lines are weaker than the even ones, as would be expected for this homonuclear
molecule.  There is also evidence for the lowest J-levels of the much
weaker (0,1) band with bandhead at 4278~\AA.

When N$_2^+$ is observed in a cometary spectrum it is
often erroneously attributed to being from the comet;
N$_2^+$ is excited in the Earth's atmosphere, especially near dusk and dawn,
and it is the telluric lines that are most often what are detected.
\citet{cocoba00} derived very tight upper limits for
N$_2^+$ for comets 122P/deVico and C/1995 O1 (Hale-Bopp). 
Other ionic species, including CO$^+$,
were observed in these cometary spectra. 
This raises the question of why we believe that the N$_2^+$
seen in R2's spectrum is cometary in nature.  The evidence is actually
quite strong.  First, the comet was observed in the middle of the night,
when we would not expect much telluric emission.  Second, other comets
observed on the same nights did not show this band at all.
Third, the band
was observed in all spectra on all three nights. Finally, the lines are
precisely at the correct wavelength for the cometary Doppler shift and  are
not coincident with the telluric restframe.  This last statement relies on
the high resolving power of the coud{\'e} spectra for certainty.

In addition to these clearly defined bands, we see evidence of the
forbidden oxygen transitions of O ($^1$D) and possibly O ($^1$S).  We also
see some additional emission lines that we have yet to identify.  However,
we can eliminate emissions due to CO$_2^+$, CH$^+$ and H$_2$O$^+$.

\section{Analysis and Implications}
Since R2 cannot be on its first passage into the inner Solar System, it is
surprising to detect two such volatile species and nothing else.  But these
two species are interesting to observe  as we expect preferentially for C to be
bound into CO or CO$_2$ and N into N$_2$ when they formed in the outer
Solar System \citep{lepr80,moetal2012,chro02}.
With our identifications
of the ions of these species, we can determine the N$_2$/CO ratio of the
ices in this comet.

From the band intensity and some physical constants, one can compute the 
column density as \[ N=I_{\nu^\prime\nu^{\prime\prime}}/g_{\nu^\prime\nu^{\prime\prime}} \]
where N is the column density, I$_{\nu^\prime\nu^{\prime\prime}}$ is the integrated band intensity and
$g_{\nu^\prime\nu^{\prime\prime}}$ is the excitation factor. From this it
follows that 
\[ \frac{\mathrm{N}_2^+}{\mathrm{CO}^+} =
        \frac{g_{\mathrm{CO}^+}}{g_{\mathrm{N}_2^+}}
        \frac{\mathrm{I}_{\mathrm{N}_2^+}} {\mathrm{I}_{\mathrm{CO}^+}} \]
yields the ratio of the column density of N$_2^+$/CO$^+$.

We measured the band intensity simply by marking a continuum and summing
all of the flux above that continuum.  For CO$^+$ we measured both ladders
of the (2,0) band separately and summed them.  For N$_2^+$, we measured the
whole P-branch and R(1) -- R(5) together and then added in the additional flux
contributions of R(6) -- R(16). 
The N$_2^+$ and CO$^+$ (2,0) bands are close in wavelength but they are
still 7 orders apart on the CCD and well off of the grating blaze. 
Thus, it is likely that the throughput is
slightly different for the two orders.  We did observe flux standards on
each night. However, these two bands are at the wavelengths of the Balmer
decrement in the A stars typically used for standards.  Thus, there are no
calibrations of this region. Instead, we used the solar spectrum 
from the daytime sky, obtained
through the same spectrograph via a ground-glass port, in
order to figure the relative flux of these orders when compared with the
atlas of \citet{kufubr84}. We determined we needed
to increase the N$_2^+$ flux by a factor of 2.0 to have
comparable throughput to the CO$^+$ order.
The excitation factor used for CO$^+$ 
was $3.55\times10^{-3}$ photons~sec$^{-1}$~mol$^{-1}$
\citep{maah86}. The excitation factor for N$_2^+$ was
$7.00\times10^{-2}$ photons~sec$^{-1}$~mol$^{-1}$ \citep{luwowa93}.
Putting together these various factors, we found that N$_2^+$/CO$^+$ = 0.15.

Converting from the quantity of the ions to the quantity of the 
neutrals is dependent on our understanding of the photodestruction
branching ratios that are not well understood. 
One possible source of CO is that some of it comes from dissociation of
CO$_2$. However, in that case we would expect to see CO$_2^+$ in our
spectra and we definitely do not detect any. Thus, we assume all the
CO$^+$ comes from the ionization of CO and convert our
measured ratio of N$_2^+$/CO$^+$ to N$_2$/CO.
\citet{wyth89} argued that one must multiply the ion ratio by two to
derive the neutral ratio, while \citet{luwowa93}
argued that no such factor is necessary. The argument of Lutz {\it et al.}
is consistent with the solar photoionization rates given in Table~3
of \citet{humu15} so we adopted this argument.
Thus, assuming that CO and N$_2$ are ionized in  proportion to the amount of
neutrals, this means that N$_2$/CO =  0.15.  

Our measured ratio is much higher than the
upper limits on this ratio found for deVico and Hale-Bopp using the same
instrument and techniques \citep{cocoba00}.  Their limits ranged from
$3\times10^{-4}$ for
deVico to $6\times10^{-5}$ for Hale-Bopp.
Indeed, it is much higher than other observations of N$_2^+$/CO$^+$,
as listed in Table III of Cochran et al.
\citet{koetal2014} measured the N$_2^+$/CO$^+$ in comet C/2002~VQ94 (LINEAR),
a comet active at $>8$\,{\sc AU}, as 0.06.  
\citet{feldfuse2015} placed a 3-$\sigma$ upper limit on N$_2$/CO of 
0.027 for comet C/2001~Q4 (NEAT) using FUSE observations.
\citet{ivetal2016} measured N$_2^+$/CO$^+$ as 0.013 for comet 
29P/Schwassmann-Wachmann~1 at 5.25\,{\sc au}, though the N$_2^+$ feature is
not well defined in these low-resolution spectra.
Using a mass spectrometer 
on the Rosetta spacecraft in orbit with comet 67P/Churyumov-Gerasimenko,
Rubin et al. (2015) measured an N$_2$/CO ratio of 
$5.7 \times 10^{-3}$. This is certainly the most robust measure of 
this ratio since it was measured {\it in situ}, though the closeness
of these species in mass makes
the measurement subject to model interpretation. 
Thus, comet R2 shows an N$_2$/CO ratio at least a
factor of 2 greater than any comet measured so far.
\citet{wiwocbet} reported on submillimeter observations
of R2, including a detection of the CO J=2--1 rotational line and a
non-detection of the HCN J=3--2 rotational transition.  They conclude
that this comet appears to be very CO-rich.
We could potentially use the O ($^1$D) 6300\AA\ lines as a proxy to determine
the abundance of water. However, as CO can also contribute photons to
this line, the derived water value would be suspect. 
Therefore, we leave detailed analysis of the O ($^1$D) 6300\AA\ line in terms
of the production of water for a future publication.

The ratio of N$_2$/CO trapped in the cometary ices is not necessarily
identical with the amount in the solar nebula.  \citet{owba95a}
used studies of deposition of gases into amorphous water ice in the
laboratory to show that ices incorporated into comets at around 50\,K would
have N$_2$/CO $\approx 0.06$ if N$_2$/CO is $\approx 1$ in the solar nebula.
Our measurement of N$_2$/CO is within a factor of 2 of their derived value,
though ours is higher.  Indeed, only some of the older
photographic data related by Arpigny (personal communication) and shown
in \citet{cocoba00} come close to the Owen and Bar-Nun prediction.

It must be remembered that the ratio of species seen in the gas phase is
not necessarily representative of the ratio of the ices in the nucleus.
However, CO and N$_2$ are not terribly reactive with other
species so chemical reactions probably do not alter this ratio much.
It also means we do not expect to see the ratio change with heliocentric
distance.
Additionally, \citet{baklko88} showed that CO and N$_2$ should be
released in the same proportion as they exist in the ices.

Owen and Bar-Nun predicted that there would be higher values of N$_2$/CO 
for dynamically new comets than for older comets.  However, as pointed out
earlier, R2 has a period of around 20,000 years and a semi-major axis of
around 1500\,{\sc AU} suggesting that it has been near the Sun prior to
this apparition.
Note that the comets with measured N$_2$/CO ratios represent a variety
of dynamical types, from
Jupiter Family comets such as 67P and 29P, to dynamically new comets
such as C/1940~R2 (Cunningham).  There is no clear trend of the
magnitude of this ratio with dynamical type.

Comet C/2016~R2 (Pan-STARRS) showed an unusual optical spectrum with strong
CO$^+$ and N$_2^+$ emissions and none of the usual neutrals seen in most
cometary spectra.  This intriguing object  showed the strongest and clearest
N$_2^+$ emissions ever detected with modern digital spectra.  Faced with
this unusual spectrum, we have alerted many members of the cometary
community and they (and we) are requesting follow-up observations with a
variety of instruments at all wavelengths in order to try to understand
this unusual comet.  

\acknowledgements
ALC was supported by NASA Grant NNX17A186G.
AJM is funded through the NASA Postdoctoral Program, administered
by the Universities Space Research Association.

\facilities{McD:0.8m, Smith (2DCoude)}

\newpage
%\appendix

\begin{center}
{\bf\large  Erratum: ``Strong CO$^+$ and N$_2^+$ Emission in Comet C/2016 R2 (Pan-STARRS)" }  \\ [5pt]
{\small
Anita L. Cochran$^1$ and Adam J. McKay$^2$ \\
$^1$McDonald Observatory, University of Texas at Austin, Austin, TX, USA \\
$^2$NASA Goddard Space Flight Center/Universities Space Research Association, 
Greenbelt, MD, USA
}
\end{center}

After publication of these observations, we discovered an error in our
spreadsheet where we summed up the band intensities.  Unfortunately, the
flux in the P branch lines  plus R(0) to R(5) got added in three times. 
As a result, we calculated a much higher N$_2^+$ band intensity than we
should have.  After correcting this error, we find that N$_2^+$/CO$^+$=
0.06.  By the arguments we included then N$_2$/CO = 0.06.  This is more in
keeping with the highest previous values detected and is consistent with
the values of Owen and Bar-Nun (1995).  However, our N$_2^+$ detection is
the cleanest detection ever because it is digital and at high spectral
resolution (see Figure 3 of original paper). 
Furthermore, we have continued monitoring this comet and still
see N$_2^+$ and CO$^+$ but not more normal neutrals.  The comet appears
extremely CO rich and depleted in many species.

\vspace{30pt}
\begin{center}
REFERENCES
\end{center}
Owen, T. \& Bar-Nun, A. 1995, Icarus, 116, 215

\end{document}